\documentclass[10pt,journal]{IEEEtran}

\newcommand{\black}[1]{{\textcolor[rgb]{0,0,0}{#1}}}

\usepackage{cite}
\usepackage{amsmath,amssymb,amsfonts}
\usepackage{algorithmic}
\usepackage{graphicx}
\usepackage{textcomp}
\usepackage{xcolor}
\usepackage{booktabs}
\usepackage{adjustbox}
\usepackage{algorithm}
\usepackage{tikz}
\usepackage{pgfplots}

\pgfplotsset{compat=1.18}

\usetikzlibrary{
  arrows.meta,
  positioning,
  fit,
  calc,
  shapes.geometric
}

\usepgfplotslibrary{groupplots}

\begin{document}

\title{Semantic Tanner-Graph Decoding for LDPC Codes}

\author{
Wenjing~Wei,
Chentao~Yue,~\IEEEmembership{Member,~IEEE},
Branka~Vucetic,~\IEEEmembership{Life Fellow,~IEEE},
and~Yonghui~Li,~\IEEEmembership{Fellow,~IEEE}%
\thanks{W. Wei, C. Yue, B. Vucetic, and Y. Li are with the School of Electrical and Computer Engineering, The University of Sydney, Sydney, NSW 2006, Australia (e-mail: \{wenjing.wei, chentao.yue, branka.vucetic, yonghui.li\}@sydney.edu.au). (\emph{Corresponding author: Chentao Yue.}) The work of Chentao Yue was supported by ARC DECRA under Grant DE250101332. Code available: https://github.com/Wenjing79/Semantic-Tanner-Graph.}%
}

\maketitle

\begin{abstract}
Conventional low-density parity-check (LDPC) belief propagation (BP)
decoding operates on a Tanner graph, 
{with} no factor describing the {source structure}.
This paper proposes a semantic Tanner graph for LDPC-coded text transmission and develops semantic-factor-aided BP
(SF-BP) \black{decoding}.
\black{The source text is tokenized into discrete tokens, which are then mapped to binary messages for channel coding.} For each token, the proposed graph introduces a semantic factor connected to the
corresponding information-bit variable nodes \black{forming the token}. A fine-tuned
language model assigns this factor a posterior distribution \black{over all possible token symbols in the alphabet at each token position.}
During SF-BP decoding, semantic-factor and LDPC parity-check messages
are \black{combined at} their shared variable nodes \black{for message passing}.
Simulation results show \black{that SF-BP significantly improves the} block error rate (BLER) and \black{Sentence-BERT similarity}
\black{(BERT: Bidirectional Encoder Representations from Transformers)}
compared with conventional BP. \black{For example, SF-BP reduces
BLER from \black{approximately $1$} to \black{$6.51\times10^{-6}$} and increases Sentence-BERT similarity
from
$0.21$ to \black{$1-1.26\times10^{-6}$} for the  (1248, 1040) LDPC code at 4 dB.}

\end{abstract}


\vspace{-0.5em}
\section{Introduction}
Reliable text transmission over noisy wireless channels requires the
receiver to recover the transmitted bit stream from distorted channel
observations \cite{10486856}. In conventional coded systems, the channel
decoder operates on the received symbols and the code constraints,
without explicitly representing the structure of the transmitted source
\cite{9955312}. Belief propagation (BP) decoding of low-density
parity-check (LDPC) codes, for example, exchanges soft information
between variable nodes and parity-check nodes on a Tanner graph
\cite{10507064}. This graph captures channel likelihoods and algebraic
parity-check constraints, but it contains no factor describing the
statistical structure of natural-language text. \black{Consequently, conventional BP does not explicitly exploit any source
structure retained in the channel-encoder input~\cite{8461983}.}

Semantic communication has attracted growing interest as a way to
exploit source structure beyond conventional bit-level reconstruction
\cite{yue2026semantic_osd}. Neural joint source-channel coding (JSCC)
learns an end-to-end mapping from source data to channel symbols and
can improve task-level or perceptual performance under noisy channels
\cite{9398576}. However, such designs often replace the
standard channel-coding interface and their performance depends on the
source modality, target task, and assumed channel model.
\black{An alternative is to retain conventional channel coding and modulation
while introducing a pretrained language model (LM) at the receiver
for semantic error control (SEC)~\cite{hao2026}.}
\black{
Recent studies  incorporate LM information into
conventional channel decoding, including LLM-Viterbi for convolutional
codes \cite{li2026llmviterbi}, \black{semantic ordered statistics decoding (Sem-OSD)}
for short block codes~\cite{yue2026semantic_osd}, and LM-assisted LDPC decoding through
bit-level soft-information feedback \cite{11498322}. More recently,
Park \emph{et al}.~\cite{park2026llmassisted} proposed a syndrome-verified
LM-assisted LDPC decoder that injects verified correction candidates
as bit-level priors for a subsequent BP \black{run}.  However, these schemes do not retain the full token posterior as a
joint factor during iterative decoding, and thus do not explicitly
exploit intra-token bit dependence through
semantic-factor message updates.}

Motivated by this, \black{we propose a semantic Tanner graph for LDPC-coded
text transmission with source compression deliberately omitted to
retain linguistic redundancy, adding one semantic factor per token
to the conventional Tanner graph.}
\black{Each semantic factor connects
to the information-bit variables forming that token and assigns
LM-derived soft weights to their possible joint configurations, while
the original LDPC parity-check factors remain unchanged.} Although we
focus on text, the construction applies to any tokenizable source for
which a probabilistic model over tokens is available.
We further develop semantic-factor-aided BP (SF-BP) on the augmented
graph. \black{Each semantic factor retains the
complete byte posterior over the corresponding eight systematic bits.
Within \black{each SF-BP run}, the factor values remain fixed, while the
factor-to-bit messages are recomputed from the evolving extrinsic
beliefs of the other bits in the byte. This allows semantic and
parity-check information to interact iteratively without additional
LM invocations. }
\black{Further LM invocations refresh the factors when an \black{SF-BP}
run ends with a nonzero syndrome.}
Simulations compare SF-BP with conventional BP, LM-Top1,
bit-marginal BP (BM-BP), \black{SV-GPT5.5~\cite{park2026llmassisted},} and DEFLATE-\black{LDPC}. BM-BP uses the same
byte posterior but removes the extrinsic-message weighting,
\black{whereas DEFLATE-\black{LDPC} combines source compression with \black{rate-adaptive LDPC coding}.}
For the $(1248,1040)$ LDPC code at
4~dB, SF-BP achieves a BLER of $6.51\times10^{-6}$, compared with
\black{approximately $1$} for BP and \black{$2.23\times10^{-5}$} for BM-BP.

\vspace{-0.5em}
\section{System Model}
\vspace{-0.8em}
\subsection{Transmission Structure}
\black{We consider single-user transmission of a tokenized natural-language
sentence over an additive white Gaussian noise (AWGN) channel using a
systematic LDPC code.}
\black{
Let $\mathcal{T}$ denote a source text block represented by a fixed
tokenizer as
$\boldsymbol{\mu}=(\mu_1,\ldots,\mu_L)$.
Each token $\mu_s$ belongs to a finite vocabulary $\mathcal{Q}$ with
$Q\triangleq|\mathcal{Q}|=2^B$ and is represented by a fixed
$B$-bit label
$\boldsymbol{\beta}(\mu_s)\in\mathbb{F}_2^B$.
Tokenization here replaces conventional source coding
\cite{qiao2025token}, where the linguistic redundancy retained in the
token stream is exploited by the receiver\footnote{\black{The fixed-length
byte representation prevents bit errors in one byte from propagating to
subsequent bytes, whereas for variable-length source codes, bit errors may
potentially affect subsequent decoded symbols~\cite{476338}.}}.
}
For a systematic $(N,K_{\rm c})$ LDPC code, we choose
$L=K_{\rm c}/B$, so that concatenating the $B$-bit token labels
directly forms the information vector
$\mathbf b\in\mathbb{F}_2^{K_{\rm c}}$.
The encoder maps $\mathbf b$ to
$\mathbf c=\operatorname{Enc}(\mathbf b)\in\mathbb{F}_2^N$ satisfying
$\mathbf H\mathbf c^{\mathsf T}=\mathbf 0$, where
$\mathbf H\in\mathbb{F}_2^{M\times N}$ is the parity-check matrix.
Because the code is systematic, each information bit $b_k$ occupies
a known codeword position $\iota(k)$.\black{\footnote{\black{
Confidentiality is provided by encryption and is outside the
scope of this work.}}}

Each coded bit is mapped to the
\black{binary phase-shift keying (BPSK)} symbol
$x_j=1-2c_j$, and the receiver observes
$y_j=x_j+w_j$, $j=1,\ldots,N$, where
$w_j\overset{\mathrm{i.i.d.}}{\sim}\mathcal{N}(0,\sigma^2)$
with $\sigma^2=N_0/2$. \black{Since the BPSK symbol energy is
$E_s=1$, we define
\(\mathrm{SNR}=1/\sigma^2=2E_s/N_0=2R E_b/N_0\),
where \(R=K_c/N\) and \(E_b\) is the energy per LDPC information bit.
Hence,
\(\mathrm{SNR}_{\rm dB}=(E_b/N_0)_{\rm dB}+10\log_{10}(2R)\).}
The receiver
forms the channel LLR
\begin{equation}
    \ell_{{\rm ch},j}
    =
    \log\frac{p(y_j\mid c_j=0)}{p(y_j\mid c_j=1)}
    =
    \frac{2y_j}{\sigma^2},
    \qquad j=1,\ldots,N,
    \label{eq:full_channel_llr}
\end{equation}
\black{which depends only on the conditional channel
likelihoods and does not include source-dependent prior information.}
The vector
$\boldsymbol{\ell}_{\rm ch}
=(\ell_{{\rm ch},1},\ldots,\ell_{{\rm ch},N})$
{serves as} the physical-layer evidence for
{BP} decoding.

\vspace{-0.5em}
\subsection{Conventional LDPC Tanner Graph and BP Receiver}

The LDPC code is represented by the Tanner graph
$\mathcal{G}_{\rm LDPC}=(\mathcal{V},\mathcal{C},\mathcal{E}_{\rm pc})$, where $\mathcal{V}=\{v_i\}_{i=1}^{N}$ denotes the code-bit variable
nodes, $\mathcal{C}=\{\black{f_a^{\rm pc}}\}_{a=1}^{M}$ denotes the
parity-check factors, and
$\mathcal{E}_{\rm pc}=\{(i,a):H_{a,i}=1\}$. \black{Let $\mathcal{A}(i)=\{a:H_{a,i}=1\}$ and
$\mathcal{I}(a)=\{i:H_{a,i}=1\}$ denote the  neighboring
check and variable index sets, respectively.} Each parity-check
factor is
\begin{equation}
    f_a^{\rm pc}\!\bigl(\mathbf{c}_{\black{\mathcal{I}(a)}}\bigr)
    =
    \mathbb{I}\!\left[
    \bigoplus\nolimits_{i\in\black{\mathcal{I}(a)}} c_i=0
    \right],
    \label{eq:parity_check_factor}
\end{equation}
where $\mathbb{I}[\cdot]$ is the indicator function.

The channel LLR at variable node $v_i$ defines the unary likelihood
factor
\vspace{-0.2em}
\begin{equation}
    \phi_i(c_i)
    \propto
    \exp\!\left[( 1-2c_i)\ell_{{\rm ch},i}/{2}\right].
    \label{eq:channel_likelihood_factor}
\end{equation}
\vspace{-0.2em}
\black{For the conventional source-unaware LDPC-BP receiver, we adopt a
uniform prior over codewords, under which the posterior factorizes as}
\begin{equation}
    p_{\rm LDPC}(\mathbf{c}\mid\mathbf{y})
    \propto
    \prod\nolimits_{i=1}^{N}\phi_i(c_i)
    \prod\nolimits_{a=1}^{M}
    f_a^{\rm pc}\!\left(\mathbf{c}_{\black{\mathcal{I}(a)}}\right).
    \label{eq:ldpc_factorization}
\end{equation}

Sum-product BP on $\mathcal{G}_{\rm LDPC}$ exchanges the
variable-to-check LLR $q_{i\rightarrow a}$ and the check-to-variable
LLR $r_{a\rightarrow i}$ along each parity-check edge. With
$r_{a\rightarrow i}^{(0)}=0$, the messages at iteration $\ell$ are


\begin{gather}
    q_{i\rightarrow a}^{(\ell)}
    =
    \ell_{{\rm ch},i}
    +
    \sum\nolimits_{a'\in\black{\mathcal{A}(i)}\setminus\{a\}}
    r_{a'\rightarrow i}^{(\ell-1)},
    \label{eq:bp_variable_to_check}\\
    r_{a\rightarrow i}^{(\ell)}
    =
    2\tanh^{-1}\!\left[
    \prod\nolimits_{i'\in\black{\mathcal{I}(a)}\setminus\{i\}}
    \tanh\!\left({q_{i'\rightarrow a}^{(\ell)}}/{2}\right)
    \right].
    \label{eq:bp_check_to_variable}
\end{gather}
The a posteriori LLR is then
\begin{equation}
    L_{\mathrm{APP},i}^{(\ell)}
    =
    \ell_{{\rm ch},i}
    +
    \sum\nolimits_{a\in\black{\mathcal{A}(i)}}
    r_{a\rightarrow i}^{(\ell)}.
    \label{eq:bp_app_llr}
\end{equation}
The hard decision is $\hat{c}_i^{(\ell)}=0$ if
$L_{\mathrm{APP},i}^{(\ell)}\geq0$ and  $1$
otherwise. \black{The conventional BP run  terminates when}  
$\mathbf{H}\hat{\mathbf{c}}^{(\ell)\mathsf T}=\mathbf{0}$; otherwise,
BP continues until the iteration limit is reached.
The factorization in \eqref{eq:ldpc_factorization} contains no source
factor, although stalled BP outputs may still retain linguistic structure.

\vspace{-0.5em}
\section{Semantic Tanner Graph Decoding}

This section first augments the LDPC Tanner graph with a  semantic-factor layer, and then introduces SF-BP. 
\vspace{-1em}
\subsection{Semantic Tanner Graph}

\black{Since the LDPC encoder is systematic, each information bit $b_k$ is
carried by a known codeword variable.}
For each source-token position $s {\in\{1,\ldots,L\}}$,
the semantic Tanner graph introduces one factor node $f_s^{\rm sem}${, attached} to the variable nodes
{carrying} the $B$ information bits of token $s$.
With $\iota(k)$ denoting the codeword position carrying
information bit $b_k$ under the systematic encoder, the semantic
neighborhood of token $s$ is
\begin{equation}
    \mathcal{B}_s
    =
    \left\{
    i_{s,k}=\iota\!\left(B(s-1)+k\right)
    \mid k=1,\ldots,B
    \right\}.
    \label{eq:semantic_neighbourhood}
\end{equation}
{Thus, the variable nodes} associated with this token are
$\mathbf{v}_{\mathcal{B}_s}=(v_{i_{s,1}},\ldots,v_{i_{s,B}})$.

Let $\mathcal{S}=\{f_s^{\rm sem}\}_{s=1}^{L}$ denote the semantic-factor
layer, whose edge set is
$\mathcal{E}_{\rm sem}=\{(i,s)\mid i\in\mathcal{B}_s,\ s=1,\ldots,L\}$.
The augmented Tanner graph is therefore
\begin{equation}
    \mathcal{G}_{\rm ST}
    =
    \left(
    \mathcal{V},
    \mathcal{C}\cup\mathcal{S},
    \mathcal{E}_{\rm pc}\cup\mathcal{E}_{\rm sem}
    \right).
    \label{eq:semantic_tanner_graph}
\end{equation}

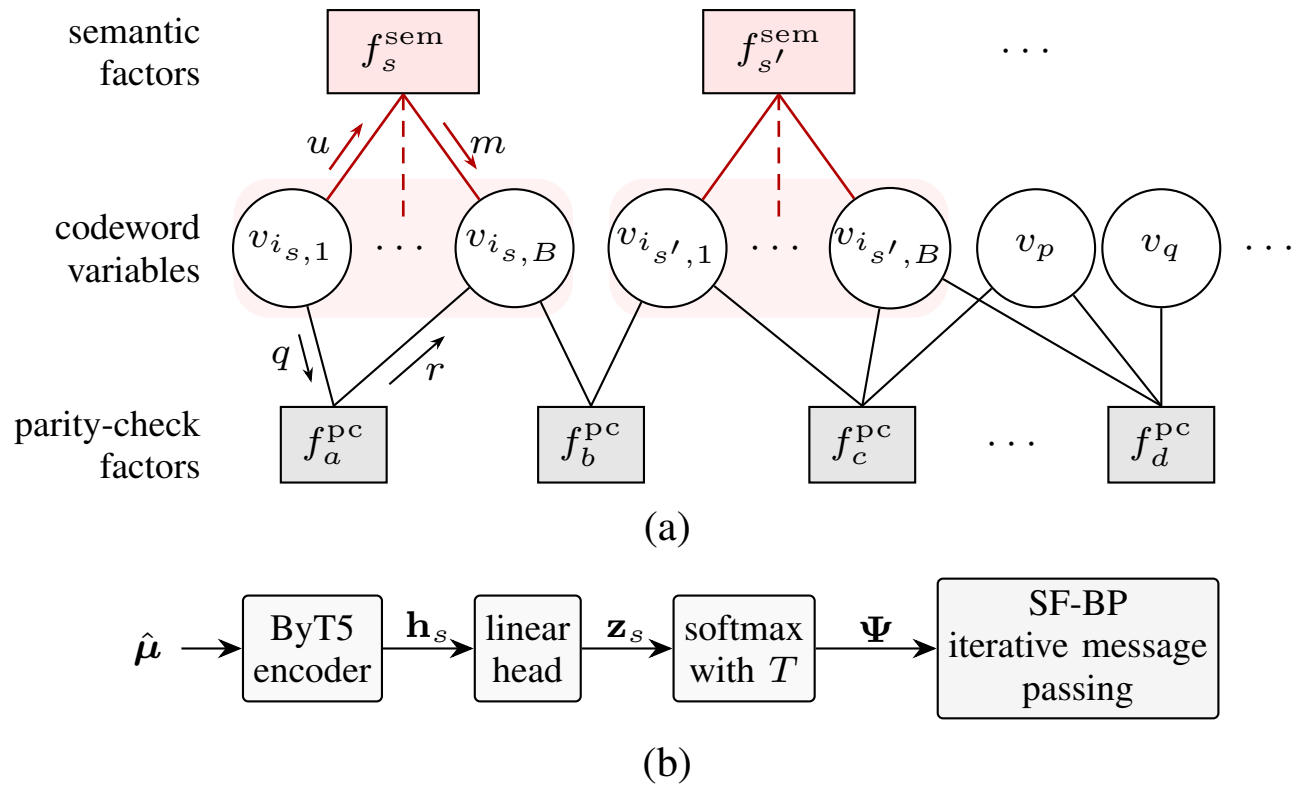
\begin{figure}[!t]
\centering

\begin{adjustbox}{max width=\linewidth}
\begin{tikzpicture}[
    x=0.92mm,
    y=1mm,
    font=\scriptsize,
    var/.style={
        circle,
        draw,
        line width=0.4pt,
        fill=white,
        minimum size=7.8mm,
        inner sep=0pt
    },
    sem/.style={
        rectangle,
        draw,
        line width=0.4pt,
        fill=red!10,
        minimum width=10mm,
        minimum height=5.5mm,
        inner sep=1pt
    },
    chk/.style={
        rectangle,
        draw,
        line width=0.4pt,
        fill=black!10,
        minimum width=7mm,
        minimum height=5mm,
        inner sep=1pt
    },
    semedge/.style={
        line width=0.5pt,
        red!70!black
    },
    pcedge/.style={
        line width=0.4pt,
        black!60!black
    },
    omit/.style={
        densely dashed
    },
    msgarr/.style={
        -{Stealth[length=1mm,width=0.7mm]}
    },
    msglab/.style={
        font=\scriptsize,
        inner sep=0.1pt
    },
    lab/.style={
        align=right,
        anchor=east,
        font=\scriptsize,
        inner sep=0pt
    }
]

\fill[red!5, rounded corners=2.4mm]
    (-0.2,6.4) rectangle (24.2,15.6);
\fill[red!5, rounded corners=2.4mm]
    (26.8,6.4) rectangle (51.2,15.6);

\node[var] (u1) at (4,11)
    {$v_{i_{s,1}}$};
\node (ud) at (12,11)
    {$\cdots$};
\node[var] (uB) at (20,11)
    {$v_{i_{s,B}}$};

\node[var] (v1) at (31,11)
    {$v_{i_{s',1}}$};
\node (vd) at (39,11)
    {$\cdots$};
\node[var] (vB) at (47,11)
    {$v_{i_{s',B}}$};

\node[var] (vp) at (57.5,11)
    {$v_p$};
\node[var] (vq) at (66.5,11)
    {$v_q$};
\node at (74.5,11)
    {$\cdots$};

\node[sem] (fs) at (12,24)
    {$f_s^{\rm sem}$};
\node[sem] (fs2) at (39,24)
    {$f_{s'}^{\rm sem}$};
\node at (57,24)
    {$\cdots$};

\node[chk] (ca) at (7,-2)
    {$f_a^{\rm pc}$};
\node[chk] (cb) at (25.5,-2)
    {$f_b^{\rm pc}$};
\node[chk] (cc) at (45,-2)
    {$f_c^{\rm pc}$};
\node at (56,-2)
    {$\cdots$};
\node[chk] (cd) at (66.5,-2)
    {$f_d^{\rm pc}$};

\foreach \v in {u1,uB}
    \draw[semedge] (fs.south) -- (\v);

\foreach \v in {v1,vB}
    \draw[semedge] (fs2.south) -- (\v);

\draw[semedge,omit] (fs.south) -- (ud.north);
\draw[semedge,omit] (fs2.south) -- (vd.north);

\foreach \v in {u1,uB}
    \draw[pcedge] (ca.north) -- (\v);

\foreach \v in {uB,v1}
    \draw[pcedge] (cb.north) -- (\v);

\foreach \v in {v1,vB,vp}
    \draw[pcedge] (cc.north) -- (\v);

\foreach \v in {vB,vp,vq}
    \draw[pcedge] (cd.north) -- (\v);


\draw[semedge,msgarr]
    (6.7,16.2) -- (9.1,19.3);
\node[msglab] at (5.9,17.8) {$u$};

\draw[semedge,msgarr]
    (14.9,19.3) -- (17.3,16.2);
\node[msglab] at (18.1,17.8) {$m$};

\draw[pcedge,msgarr]
    (4.5,5.3) -- (5.4,2.1);
\node[msglab] at (3.2,3.7) {$q$};

\draw[pcedge,msgarr]
    (11.0,2.0) -- (15.0,5.2);
\node[msglab] at (14.3,2.8) {$r$};

\node[lab] at (-2.5,24)
    {semantic\\factors};
\node[lab] at (-2.5,11)
    {codeword\\variables};
\node[lab] at (-2.5,-2)
    {parity-check\\factors};

\end{tikzpicture}
\end{adjustbox}\\[-0.5mm]
{\footnotesize (a)}\\[2mm]

\begin{adjustbox}{max width=\linewidth}
\begin{tikzpicture}[
    font=\scriptsize,
    blk/.style={
        rectangle,
        draw,
        rounded corners=1pt,
        align=center,
        minimum height=7mm,
        inner sep=2pt,
        fill=gray!5
    },
    arr/.style={
        -{Stealth[length=1.6mm,width=1.3mm]},
        thin
    },
    lab/.style={
        font=\scriptsize,
        inner sep=1pt
    }
]

\node (in)
    {$\hat{\boldsymbol{\mu}}$};

\node[blk, right=4mm of in] (enc)
    {ByT5\\encoder};

\node[blk, right=6mm of enc] (head)
    {linear\\head};

\node[blk, right=6mm of head] (sm)
    {softmax\\with $T$};

\node[blk, right=8mm of sm, fill=gray!8] (sf)
    {SF-BP\\iterative message\\passing};

\draw[arr] (in) -- (enc);

\draw[arr] (enc) --
    node[lab, above] {$\mathbf{h}_s$}
    (head);

\draw[arr] (head) --
    node[lab, above] {$\mathbf{z}_s$}
    (sm);

\draw[arr] (sm) --
    node[lab, above] {$\boldsymbol{\Psi}$}
    (sf);

\end{tikzpicture}
\end{adjustbox}\\[-0.5mm]
{\footnotesize (b)}

\caption{\black{
(a) Parametrized partial semantic Tanner graph for two representative
source tokens. (b) Construction of the semantic factors from the
current decoded byte sequence.
}}
\label{fig:semantic-tanner-graph}
\vspace{-1em}
\end{figure}

\black{
Fig.~\ref{fig:semantic-tanner-graph}\black{(a)} illustrates a representative
portion of $\mathcal{G}_{\rm ST}$ for two token positions $s$ and
$s'$. Each semantic factor $f_u^{\rm sem}$, $u\in\{s,s'\}$,
connects to the $B$ systematic-bit variables associated with token
$u$. The nodes $v_p$ and $v_q$, with $p,q\in\mathcal{P}$, represent
parity-bit variables, where $\mathcal{P}$ denotes the set of parity-bit
positions. Representative parity-check factors are indexed by
$a,b,c,d\in\{1,\ldots,M\}$, with an edge between $v_i$ and
$f_a^{\rm pc}$ iff $H_{a,i}=1$. \black{
Only selected nodes and edges are shown; ellipses denote omitted
nodes, and dashed red edges represent connections to omitted
variables within each token.
}
}

\vspace{-0.5em}
\subsection{Semantic Factor}
\black{
In this work, each semantic factor is parameterized by ByT5,
a pre-trained byte-level Transformer \cite{xue2022byt5}.
We use the ByT5-small encoder and apply a shared linear layer to the encoder representation at each byte position, producing logits over the 256 possible byte values.
Given the current decoded byte sequence
$\hat{\boldsymbol{\mu}}
=(\hat{\mu}_1,\ldots,\hat{\mu}_L)$,
the encoder\black{, with $d$ hidden features per position,}
produces a contextual representation
$\mathbf{h}_s\in\mathbb{R}^{d}$ at each byte position $s$,
which depends on the whole $\hat{\boldsymbol{\mu}}$
rather than on $\hat{\mu}_s$ alone.
\black{Here, $Q=256$ and $B=8$.}
The output head \black{with weight matrix
$\mathbf{W}\in\mathbb{R}^{Q\times d}$ and bias vector
$\mathbf{a}\in\mathbb{R}^{Q}$} maps $\mathbf{h}_s$ to logits
$\mathbf{z}_s=\mathbf{W}\mathbf{h}_s+\mathbf{a}$,
where $z_s(\xi)$ is the score assigned to candidate byte
$\xi\in\mathcal{Q}$\black{, larger for bytes that better fit the context}. The corresponding byte distribution is}
\begin{equation}
p_s(\xi\mid\hat{\boldsymbol{\mu}})
=
\frac{\exp\!\left(z_s(\xi)/T\right)}
{\displaystyle\sum\nolimits_{\xi'\in\mathcal{Q}}
\exp\!\left(z_s(\xi')/T\right)},
\label{eq:token_posterior}
\end{equation}
where \black{a smaller \black{softmax temperature}
$T>0$ concentrates $p_s$ on fewer candidate bytes}.

Each byte $\xi$ maps bijectively to a $B$-bit label
$\boldsymbol{\beta}(\xi)$, so the byte distribution defines a
joint semantic factor over the associated systematic bits\black{, rather
than separate bit priors}:
$\psi_s\!\left(\boldsymbol{\beta}(\xi)\right)
\triangleq
p_s(\xi\mid\hat{\boldsymbol{\mu}})$.
The complete set of semantic factors is
$\boldsymbol{\Psi}=\{\psi_s(\cdot)\}_{s=1}^{L}$.
\black{Fig.~\ref{fig:semantic-tanner-graph}(b) summarizes this construction.}

ByT5 is fine-tuned on \black{pairs of BP-decoded sequences $\hat{\boldsymbol{\mu}}$ (input) and
clean sequences $\boldsymbol{\mu}$ (target) from LDPC-coded AWGN
transmissions, with the \black{cross-entropy (CE)} loss}
\begin{equation}
\mathcal{L}_{\mathrm{CE}}
=
-\frac{1}{L}
\sum\nolimits_{s=1}^{L}
\log p_s(\mu_s\mid\hat{\boldsymbol{\mu}}),
\end{equation}
where $p_s$ is evaluated with $T=1$.

\vspace{-0.5em}
\subsection{Belief Propagation on the Semantic Tanner Graph}

{SF-BP} on $\mathcal{G}_{\rm ST}$ exchanges the
conventional parity-edge messages $q_{i\rightarrow a}$ and
$r_{a\rightarrow i}$ together with the semantic-edge messages
$u_{i\rightarrow s}$ and $m_{s\rightarrow i}$.
\black{
The overall SF-BP procedure is an approximate iterative decoder,
rather than exact inference on a single fixed factorization of
\(p(\mathbf c\mid\mathbf y)\).}
This
subsection describes the message updates for a single SF-BP run in
which the factor values $\boldsymbol{\Psi}$ are available.

After the LDPC check-node update in \eqref{eq:bp_check_to_variable},
variable node $v_{i_{s,k}}$ sends semantic factor $s$ the extrinsic LLR
\begin{equation}
    u_{i_{s,k}\rightarrow s}^{(\ell)}
    =
    \ell_{{\rm ch},i_{s,k}}
    +
    \sum\nolimits_{a\in\black{\mathcal{A}(i_{s,k})}}
    r_{a\rightarrow i_{s,k}}^{(\ell)}.
    \label{eq:variable_to_semantic}
\end{equation}
This message contains the channel evidence and the incoming
parity-check messages, excluding the semantic message from factor $s$.
It induces the binary belief
\begin{equation}
    \eta_{i_{s,k}\rightarrow s}^{(\ell)}(x)
    \propto
    \exp\!\left[
    \frac{1-2x}{2}
    u_{i_{s,k}\rightarrow s}^{(\ell)}
    \right],
    \qquad x\in\{0,1\}.
    \label{eq:variable_semantic_belief}
\end{equation}

\black{
Let \(\nu_{s\rightarrow i_{s,k}}^{(\ell)}(x)\) denote the
probability-domain message from \(f_s^{\rm sem}\) to variable node
\(v_{i_{s,k}}\) for the hypothesis \(c_{i_{s,k}}=x\),
\(x\in\{0,1\}\). The sum-product factor-to-variable rule fixes the target bit to \(x\)
and sums over the remaining \(B-1\) bits:
\begin{equation}
\nu_{s\rightarrow i_{s,k}}^{(\ell)}(x)
\propto
\sum_{\xi\in\mathcal Q:\,\beta_k(\xi)=x}
\psi_s(\xi)
\prod_{\substack{j=1\\j\neq k}}^{B}
\eta_{i_{s,j}\rightarrow s}^{(\ell)}
\bigl(\beta_j(\xi)\bigr).
\end{equation}
Define the extrinsic weight for candidate \(\xi\) as
\begin{equation}
\omega_{s,k}^{(\ell)}(\xi)
=
\prod\nolimits_{\substack{j=1\\j\neq k}}^{B}
\eta_{i_{s,j}\rightarrow s}^{(\ell)}
\bigl(\beta_j(\xi)\bigr),
\qquad \xi\in\mathcal Q.
\label{eq:extrinsic_weight}
\end{equation}
Taking the log-ratio of the two outgoing message values gives
\begin{equation}
m_{s\rightarrow i_{s,k}}^{(\ell)}
\!=\!\log\!
\frac{\nu_{s\rightarrow i_{s,k}}^{(\ell)}(0)}
     {\nu_{s\rightarrow i_{s,k}}^{(\ell)}(1)}
\!=\!\log\!
\frac{
\displaystyle\sum_{\xi:\,\beta_k(\xi)=0}
\psi_s(\xi)\omega_{s,k}^{(\ell)}(\xi)
}{
\displaystyle\sum_{\xi:\,\beta_k(\xi)=1}
\psi_s(\xi)\omega_{s,k}^{(\ell)}(\xi)
}.
\label{eq:semantic_to_variable}
\end{equation}
Each candidate is
softly weighted by the semantic factor and the extrinsic beliefs
of the other \(B-1\) bits. Equation~\eqref{eq:semantic_to_variable} is therefore the
standard sum-product factor-to-variable update expressed in the
LLR domain.
}
\black{The semantic factor imposes an LM-derived soft constraint
rather than a deterministic XOR parity constraint.}

For an information-bit variable $i=i_{s,k}$, define
$m_{{\rm sem},i}^{(\ell)}=m_{s\rightarrow i_{s,k}}^{(\ell)}$; for
{parity-bit
variables}, set $m_{{\rm sem},i}^{(\ell)}=0$. The variable-node updates
become
\begin{align}
    q_{i\rightarrow a}^{(\ell+1)}
    &=
    \ell_{{\rm ch},i}
    +
    \sum\nolimits_{a'\in\black{\mathcal{A}(i)}\setminus\{a\}}
    r_{a'\rightarrow i}^{(\ell)}
    +
    \alpha m_{{\rm sem},i}^{(\ell)},
    \label{eq:sf_variable_to_check}\\
    L_{\mathrm{APP},i}^{(\ell)}
    &=
    \ell_{{\rm ch},i}
    +
    \sum\nolimits_{a\in\black{\mathcal{A}(i)}}
    r_{a\rightarrow i}^{(\ell)}
    +
    \alpha m_{{\rm sem},i}^{(\ell)},
    \label{eq:sf_app_llr}
\end{align}
{which generalize \eqref{eq:bp_variable_to_check}  and \eqref{eq:bp_app_llr}
by adding the scaled semantic message
$\alpha m_{{\rm sem},i}^{(\ell)}$.} \black{Here, $\alpha>0$ controls the semantic-message weight to mitigate LM
overconfidence.}

\vspace{-0.5em}
\subsection{SF-BP Decoding Schedule}

\black{
A \emph{BP iteration} $\ell$ is one round of message updates;
a \emph{BP run} comprises up to $I_{\max}$ such iterations
from one initialization.
An \emph{LM invocation} $t$ is one ByT5 forward pass on
$\hat{\boldsymbol{\mu}}^{(t)}$, yielding factor values
$\boldsymbol{\Psi}^{(t)}$ that remain fixed throughout the
subsequent SF-BP run.
}
\black{Regardless of the  syndrome of the initial BP run, its information-bit
decisions form $\hat{\boldsymbol{\mu}}^{(1)}$, and ByT5 constructs
$\boldsymbol{\Psi}^{(1)}$.
Every block thus undergoes at least one SF-BP run.
Each SF-BP run is initialized from the original channel evidence as}
    $q_{i\rightarrow a}^{(1)}
    =
    \ell_{{\rm ch},i},
    r_{a\rightarrow i}^{(0)}
    =
    0,
    m_{{\rm sem},i}^{(0)}
    =
    0.
   $
\black{Each SF-BP run stops at a zero syndrome or after
$I_{\max}$ iterations.}

\makeatletter
\let\savedfstalgorithm\fst@algorithm
\def\fst@algorithm{%
  \savedfstalgorithm
  \def\@fs@post{\kern2pt\hrule\relax\kern-10pt}%
}
\makeatother

\begin{algorithm}[t]
\small
\caption{{SF-BP Decoding Schedule}}
\label{alg:sfbp_receiver}
\begin{algorithmic}[1]
\REQUIRE  \black{Channel LLRs $\boldsymbol{\ell}_{\rm ch}$,
LM-invocation budget $N_{\rm sem}\geq1$,
and BP iteration limit $I_{\max}$ per run}
\ENSURE Codeword estimate $\hat{\mathbf{c}}_{\rm cur}$ and decoding
status

\STATE Run conventional {BP}  using
$\boldsymbol{\ell}_{\rm ch}$ to obtain
$\hat{\mathbf{c}}_{\rm cur}$.

\FOR{$t=1$ to $N_{\rm sem}$}
    \STATE Convert the current information-bit decisions into
    $\hat{\boldsymbol{\mu}}^{(t)}$.
    \STATE \black{Perform one LM invocation to construct
$\boldsymbol{\Psi}^{(t)}$}.
    \STATE \black{Initialize messages from the original channel evidence.}

    \FOR{$\ell=1$ to $I_{\max}$}
        \STATE Update $r_{a\rightarrow i}^{(\ell)}$ from
        $q_{i\rightarrow a}^{(\ell)}$ {via \eqref{eq:bp_check_to_variable}}.
        \STATE Update $u_{i\rightarrow s}^{(\ell)}$ and
        $m_{s\rightarrow i}^{(\ell)}$ {via \eqref{eq:variable_to_semantic}} and \eqref{eq:semantic_to_variable}.
        \STATE Form $\mathbf{L}_{\mathrm{APP},i}^{(\ell)}${ via \eqref{eq:sf_app_llr}} and
        $q_{i\rightarrow a}^{(\ell+1)}${ via \eqref{eq:sf_variable_to_check}}.
        \STATE Obtain $\hat{\mathbf{c}}_{\rm cur}$ from $\mathbf{L}_{\mathrm{APP},i}^{(\ell)}$.
        \IF{$\mathbf{H}\hat{\mathbf{c}}_{\rm cur}^{\mathsf T}
        =\mathbf{0}$}
            \STATE \textbf{return}
            $(\hat{\mathbf{c}}_{\rm cur},\mathrm{SUCCESS})$
        \ENDIF
    \ENDFOR
\ENDFOR

\makeatletter
\let\fst@algorithm\savedfstalgorithm
\makeatother

\STATE \textbf{return}
$(\hat{\mathbf{c}}_{\rm cur},\mathrm{FAILURE})$
\end{algorithmic}

\end{algorithm}

\black{If the $t$-th SF-BP run ends with a nonzero syndrome and
$t<N_{\rm sem}$, its latest information-bit decisions form
$\hat{\boldsymbol{\mu}}^{(t+1)}$.
The next LM invocation constructs
$\boldsymbol{\Psi}^{(t+1)}$, and the messages are
reinitialized.
This process continues until the syndrome check is satisfied
or the LM-invocation budget $N_{\rm sem}$ is exhausted.
Algorithm~\ref{alg:sfbp_receiver} summarizes the complete receiver.}

\vspace{-0.6em}
\section{Simulation Results}

\subsection{Simulation Setup}

\subsubsection{Dataset and Training}
\black{
For the main \((1248,1040)\) configuration, Wikipedia sentences
containing 130--137 ASCII bytes are truncated to 130-byte blocks,
deduplicated, and split approximately $80/10/10$ into source-disjoint
training/validation/test sets of 185,418, 23,183, and 23,133 blocks.
Training pairs are generated through the corresponding
LDPC-coded AWGN system \black{at $2$--$4$~dB in $0.5$-dB steps}.
The  \((256,128)\) configuration uses a separate dataset
and checkpoint, with training pairs generated from 16-byte
blocks at \(0\)--\(2.5\)~dB in \(0.5\)-dB steps.
ByT5-small  is
fine-tuned using AdamW with learning rate \(2\times10^{-4}\), cosine
scheduling, and 1,000 warm-up steps, and the
checkpoint with the lowest validation cross-entropy is retained.
}

\subsubsection{Channel Codes and Receiver Parameters}
The main experiments use the \((1248,1040)\) rate-\(5/6\) \black{WiMAX} LDPC code,
with auxiliary results for the \((256,128)\) rate-\(1/2\)
\black{CCSDS} code. \black{Each  BP run uses at most
$I_{\max}=50$ BP iterations.} 
For each configuration, $T$ minimizes validation byte-level
negative log-likelihood, and $\alpha$ is then selected by
validation BLER. The fixed test settings are
$(T^\star,\alpha^\star)=(1.3,0.8)$ for rate $5/6$ and
$(1.3,0.6)$ for rate $1/2$.

\subsubsection{Benchmarks and Metrics}
SF-BP is compared with \black{five} baseline receivers.
\black{We define LM-Top1 and BM-BP as controlled baselines using
the same fine-tuned ByT5 model as SF-BP.}
\emph{LDPC-BP} applies conventional BP
\black{to LDPC-coded text without source compression}.
\emph{LM-Top1} replaces each decoded token after BP failure with the
ByT5 argmax
\(\arg\max_{\xi} p_s(\xi\mid\hat{\boldsymbol{\mu}})\),
without further BP iterations.
\emph{BM-BP}, \black{conceptually related to the bit-level
soft-information feedback in~\cite{11498322},} sets
\(\omega_{s,k}^{(\ell)}(\xi)=1\) in
\eqref{eq:semantic_to_variable} to form fixed
\(\alpha\)-scaled bit-marginal LLRs under the same decoding schedule.
\black{\emph{SV-GPT5.5} follows the syndrome-verified decoder of~\cite{park2026llmassisted},
replacing its GPT-3.5-turbo corrector with GPT-5.5 and injecting the
verified corrections as fixed-magnitude bit-level priors into a
second BP run.}
\emph{DEFLATE-LDPC} \black{independently compresses each
130-byte block to $K_{\rm comp}$ bits using raw DEFLATE
(Lempel--Ziv and Huffman coding), followed by rate-matched
\black{5G LDPC coding} with BP decoding.
The budget is 1248 BPSK symbols at the same symbol energy,
yielding an effective rate $K_{\rm comp}/1248$ (mean $0.650$).
$K_{\rm comp}$ is assumed known without signaling overhead.}

\vspace{-0.8em}
\subsection{Complexity Analysis}

Let $E$ denote the number of LDPC Tanner-graph edges.
\black{
For fixed encoder depth and internal widths proportional to
the hidden dimension $d$, one LM invocation has complexity
$C_{\rm LM}=\mathcal{O}(L^2d+Ld^2+LdQ)$.
The terms correspond to pairwise attention, encoder projections
and feed-forward layers, and the byte-output head, respectively.
Table~\ref{tab:receiver_complexity} separates the per-block
worst-case cost into LDPC message passing, additional semantic
operations, and LM inference.
}

\black{
BM-BP and SF-BP use one initial BP run and
one to $N_{\rm sem}$  \black{subsequent BP} runs.
BM-BP computes bit-marginal LLRs once per LM invocation
($\mathcal{O}(BLQ)$) and holds them fixed within the BP run;
$\mathcal{O}(BL)$ per BP iteration is counted for prior
incorporation.
SF-BP recomputes semantic factor-to-variable messages
at each BP iteration ($\mathcal{O}(BLQ)$).
LM-Top1 uses at most one LM invocation and
$\mathcal{O}(LQ)$ operations for byte-wise argmax.
Syndrome-based early termination can reduce the average cost.
}

\begin{table}[t]
\centering
\caption{Per-Block Worst-Case Receiver Complexity}
\label{tab:receiver_complexity}
\setlength{\tabcolsep}{3pt}
\renewcommand{\arraystretch}{1.1}
\resizebox{\columnwidth}{!}{%
\begin{tabular}{lccc}
\toprule
Receiver
& \shortstack{LDPC\\message passing}
& \black{\shortstack{Additional\\semantic operations}}
& \shortstack{LM\\inference} \\
\midrule
LDPC-BP
& $\mathcal{O}(I_{\max}E)$
& ---
& --- \\
LM-Top1
& $\mathcal{O}(I_{\max}E)$
& $\mathcal{O}(LQ)$
& \black{$C_{\rm LM}$} \\
BM-BP
& $\mathcal{O}((1+N_{\rm sem})I_{\max}E)$
& \black{$\mathcal{O}(N_{\rm sem}BL(Q+I_{\max}))$}
& \black{$N_{\rm sem}C_{\rm LM}$} \\
SF-BP (proposed)
& $\mathcal{O}((1+N_{\rm sem})I_{\max}E)$
& $\mathcal{O}(N_{\rm sem}I_{\max}BLQ)$
& \black{$N_{\rm sem}C_{\rm LM}$} \\
\bottomrule
\end{tabular}%
}
\end{table}

\subsection{Performance Evaluation}

\begin{figure}[t]
\centering
\begin{tikzpicture}
\begin{groupplot}[
    group style={
        group size=2 by 1,
        horizontal sep=0.35cm
    },
    ymode=log,
    width=0.58\linewidth,
    height=3.8cm,
    xlabel={SNR (dB)},
    ylabel={BLER},
    ylabel shift = -5pt,
    ymin=1e-7,
    ymax=1.1,
    tick label style={font=\scriptsize},
    label style={font=\small},
    xmajorgrids,
    ymajorgrids,
    xminorgrids,
    minor grid style={solid,gray!18},
    major grid style={solid,gray!30},
    legend image post style={scale=0.75}
]

\nextgroupplot[
    xmin=2,
    xmax=6,
    xtick={2,3,4,5,6},
    minor xtick={2.5,3.5,4.5,5.5},
    legend to name=fusionlegend,
    legend columns=3,
    legend style={
        legend cell align=left,
        draw=gray!55,
        font=\scriptsize,
        inner sep=1.2pt,
        column sep=5pt,
        row sep=-1pt,
        fill=white,
        fill opacity=0.85,
        text opacity=1
    },
    after end axis/.code={
        \node[font=\small, anchor=north]
        at (axis description cs:0.5,-0.3) {(a) $R=5/6$};
    }
]

\addplot[
    mark=square*,
    mark size=2pt,
    color=gray!75!black,
    thick
] coordinates {
(2,1)
(2.5,1)
(3,1)
(3.5,1)
(4,0.999966168388)
(4.5,0.9811981)
(5,0.6281467)
(5.5,0.100831704029)
(6,0.00360282788412)
};
\addlegendentry{LDPC-BP}

\addplot[
    mark=star,
    mark size=2.0pt,
    color=magenta!85!black,
    thick
] coordinates {
(2,0.98046875)
(2.5,0.953857421875)
(3,0.916259765625)
(3.5,0.880859375)
(4,0.799072265625)
(4.5,0.690186)
(5,0.371582)
(5.5,0.0512695)
(6,0.00153665)
};
\addlegendentry{LM-Top1}

\addplot[
    mark=pentagon*,
    mark size=2.0pt,
    color=orange!85!black,
    thick
] coordinates {

(2,1)
(2.5,1)
(3,1)
(3.5,0.9433962)
(4,0.564972)
(4.5,0.15)
(5,0.019574)
(5.5,0.00214)
(6,0.0002339227)
};
\addlegendentry{SV-GPT5.5~\cite{park2026llmassisted}}

\addplot[
    mark=triangle*,
    mark size=1.8pt,
    color=green!80!black,
    thick
] coordinates {
(2,0.843881856540)
(2.5,0.539083557951)
(3,0.240384615385)
(3.5,0.0484613520717)
(4,0.00280104198762)
(4.5,8.60281656214e-5)
(5,3.3e-6)
(5.5,1e-7)
};
\addlegendentry{DEFLATE-LDPC}

\addplot[
    mark=*,
    mark size=1.8pt,
    color=black!80!black,
    thick
] coordinates {
(2,0.027099609375)
(2.5,0.00417073567708)
(3,0.000478800060680)
(3.5,8.17891541039e-5)
(4,2.23469679634e-5)
(4.5,6.44851096144e-6)
(5,2.2e-6)
(5.5,1e-6)
(6,5.38151033594e-7)
};
\addlegendentry{BM-BP}

\addplot[
    mark=diamond*,
    mark size=2.0pt,
    color=red!85!black,
    thick
] coordinates {
(2,0.00905539772727)
(2.5,0.000934709821429)
(3,0.000113159038239)
(3.5,2.44008792303e-5)
(4,6.50607928048e-6)
(4.5,2.9e-6)
(5,1.4e-6)
(5.5,8.24546960918e-7)
(6,4.54667140005e-7)
};

\addlegendentry{SF-BP}

\nextgroupplot[
    xmin=0.5,
    xmax=2.5,
    xtick={1,1.5,2,2.5},
    minor xtick={0.5},
    ylabel={},
    yticklabels={,,,},
    after end axis/.code={
        \node[font=\small, anchor=north]
        at (axis description cs:0.5,-0.3) {(b) $R=1/2$};
    }
]

\addplot[
    mark=square*,
    mark size=1.7pt,
    color=gray!75!black,
    thick
] coordinates {
(0.5,0.963245738636364)
(1,0.8314453125)
(1.5,0.570689510723861)
(2,0.27517)
(2.5,0.08589)
};

\addplot[
    mark=star,
    mark size=2.0pt,
    color=magenta!85!black,
    thick
] coordinates {
(0.5, 0.689453)
(1,0.587891)
(1.5,0.387695)
(2,0.168945)
(2.5,0.0535482)
};

\addplot[
    mark=*,
    mark size=1.8pt,
    color=black!80!black,
    thick
] coordinates {
(0.5,0.074462890625)
(1,0.0233609068627451)
(1.5,0.00642503415300546)
(2,0.00112)
(2.5,0.00014)
};

\addplot[
    mark=diamond*,
    mark size=2.0pt,
    color=red!85!black,
    thick
] coordinates {
(0.5,0.0569957386363636)
(1,0.0146484375)
(1.5,0.00314175603217158)
(2,0.00056)
(2.5,0.00009)
};

\end{groupplot}

\path (group c1r1.north) -- (group c2r1.north)
    node[midway, yshift=6mm] {\ref{fusionlegend}};

\end{tikzpicture}
\caption{ BLER comparison of semantic decoding strategies for
(a) \black{the $(1248,1040)$} rate-$5/6$ and (b) \black{the $(256,128)$} rate-$1/2$  LDPC codes.}
\vspace{-1em}
\label{fig:bler_fusion_comparison}
\end{figure}
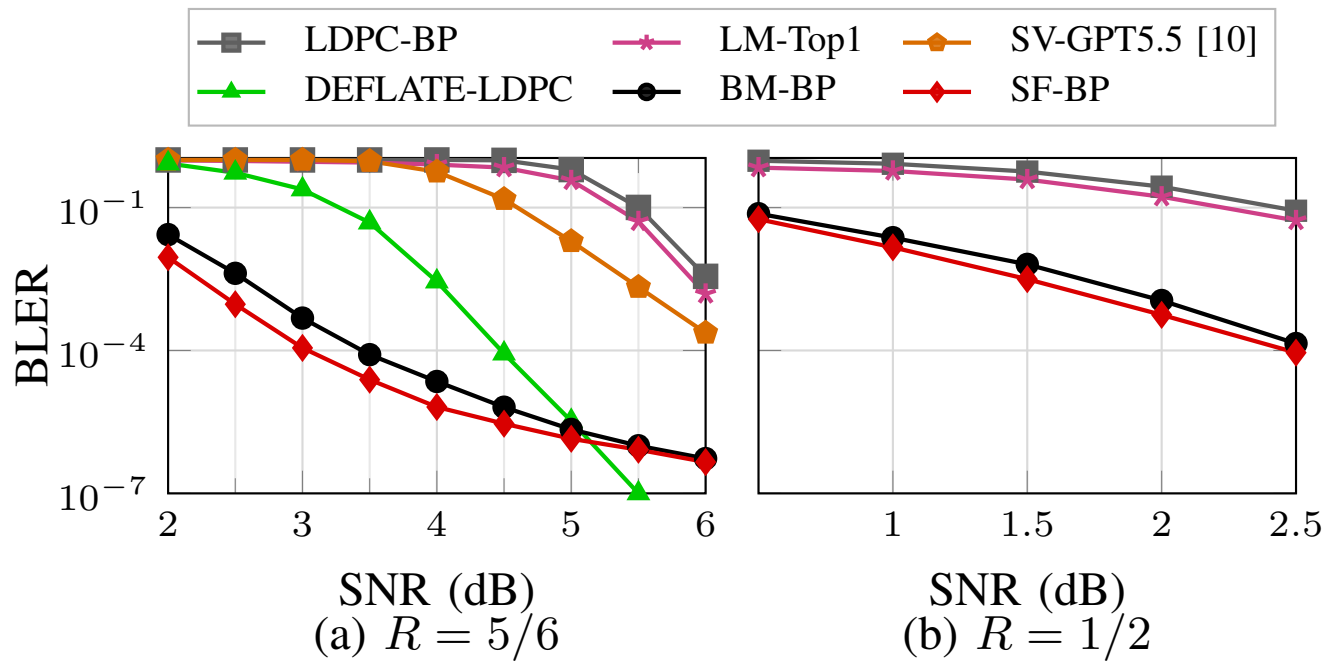

\subsubsection{\black{BLER Performance}}

Fig.~\ref{fig:bler_fusion_comparison}  compares the receivers for the
rate-$5/6$ and rate-$1/2$ configurations.
\black{LM-Top1} provides only limited improvement, as its hard token
corrections are not fed back into BP. Both BM-BP and SF-BP benefit from ByT5-assisted BP, while SF-BP
consistently improves upon BM-BP by retaining the joint byte posterior
and updating its factor messages using the evolving extrinsic beliefs.
\black{For rate $5/6$ at $4$~dB, SV-GPT5.5 gives
$5.65\times10^{-1}$, whereas SF-BP achieves
$2.09\times10^{-5}$ with one LM invocation
(\black{Section~\ref{sec:lm_budget}}).
SF-BP provides approximately $1.43$~dB gain over DEFLATE-LDPC
at BLER $10^{-4}$,
but DEFLATE-LDPC achieves lower BLER at $5.5$~dB.
For rate $5/6$, the reduced high-SNR BLER slope is
associated with imperfect LM estimates of the transmitted
source. The byte posterior learned from training data
can favor incorrect candidates, producing misleading
semantic feedback that limits further error correction
and contributes to the observed error-floor behavior.
Fig.~\ref{fig:bler_fusion_comparison}(b)  \black{compares BP with the receivers sharing the same fine-tuned ByT5, thus excluding SV-GPT5.5}.
DEFLATE-LDPC is omitted because raw DEFLATE \black{yields no compression gain on} these 16-byte blocks.}

\begin{figure}[t]
\centering
\begin{tikzpicture}
\begin{axis}[
    width=0.86\linewidth,
    height=3.3cm,
    xlabel={SNR (dB)},
    ylabel style={align=center},
    ylabel={Sentence-BERT\\similarity},
    xmin=2,
    xmax=6,
    ymin=0,
    ymax=1.05,
    xtick={2,3,4,5,6},
    ytick={0,0.2,0.4,0.6,0.8,1},
    tick label style={font=\footnotesize},
    label style={font=\small},
    xmajorgrids,
    ymajorgrids,
    major grid style={solid,gray!30},
    legend image post style={scale=0.85},
    legend style={
        at={(0.5,1.03)},
        anchor=south,
        legend columns=3,
        legend cell align=left,
        draw=gray!55,
        font=\tiny,
        inner sep=1.2pt,
        row sep=-2pt,
        fill=white,
        fill opacity=0.82,
        text opacity=1,
        /tikz/every even column/.append style={column sep=2pt}
    }
]
\addplot[
    mark=*,
    mark size=2.2pt,
    color=gray!75!black,
    thick
] coordinates {
(2,0.0705546724)
(3,0.1148251804)
(4,0.2145980966)
(5,0.6539051979)
(6,0.9985942365)
};
\addlegendentry{LDPC-BP}

\addplot[
    mark=triangle*,
    mark size=2.2pt,
    color=green!85!black,
    thick
] coordinates {

(2,0.174404807373)
(2.5,0.475500599448)
(3,0.765460671687)
(3.5,0.953001075816)
(4,0.997281332925)
(4.5,0.999916762403)
(5,0.999996827714)
(5.5,0.999999892039)
(6,1)
};
\addlegendentry{DEFLATE-LDPC}

\addplot[
    mark=square*,
    mark size=2.2pt,
    color=black!85!black,
    thick
] coordinates {
(2,0.7388305306)
(3,0.8258627576)
(4,0.8881764943)
(5,0.9639406199)
(6, 0.9998538464)
};
\addlegendentry{LM-Top1}

\addplot[
    mark=square*,
    mark size=2pt,
    color=black!85!black,
    thick
] coordinates {
(2,0.990088212691)
(3,0.999846690873)
(4,0.999993667598)
(5,0.999999846576)
(6,0.999999970772)
};
\addlegendentry{BM-BP}

\addplot[
    mark=diamond*,
    mark size=2.2pt,
    color=red!85!black,
    thick
] coordinates {
(2,0.996603378104)
(3,0.999968513965)
(4,0.999998738755)
(5,0.999999909549)
(6,0.999999979439)
};
\addlegendentry{SF-BP}

\end{axis}
\end{tikzpicture}
\caption{ Sentence-BERT similarity versus SNR for different decoding strategies with
\black{the $(1248,1040)$} rate-$5/6$  LDPC code.}
\vspace{-1em}
\label{fig:text_quality}
\end{figure}
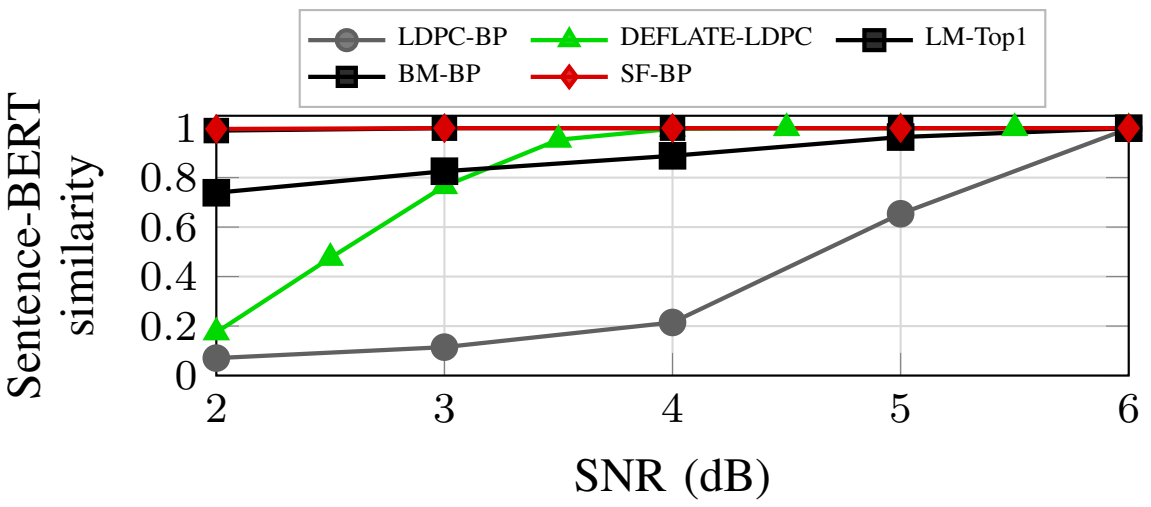

\subsubsection{Text-Recovery Quality}

Fig.~\ref{fig:text_quality} compares the average Sentence-BERT cosine
similarity for the rate-$5/6$ code. LM-Top1 improves substantially over
LDPC-BP by exploiting linguistic context.
\black{DEFLATE-LDPC improves sharply with SNR, increasing from
0.174 at 2~dB to 0.953 at 3.5~dB and approaching unity at
higher SNRs.}
BM-BP and SF-BP achieve near-unit similarity over most of the range,
with SF-BP highest at low-to-moderate SNRs. The trends are consistent
with the BLER results in Fig.~\ref{fig:bler_fusion_comparison}.
\black{The exact block-recovery rate is $1-\mathrm{BLER}$,
with undetected errors included in BLER.
A zero syndrome does not guarantee correct recovery.}

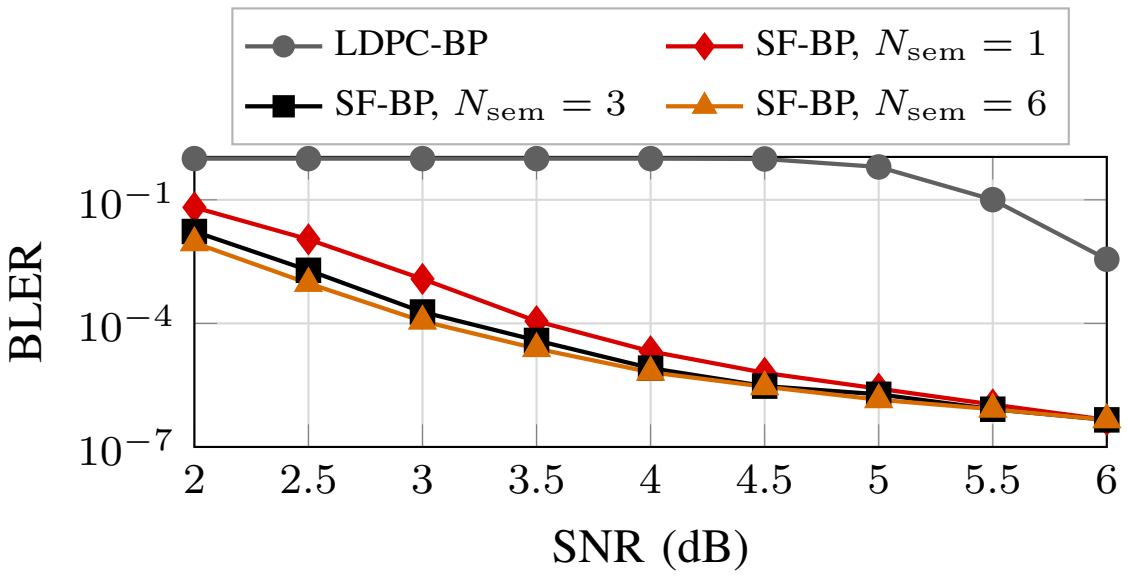
\begin{figure}[t]
\centering
\begin{tikzpicture}
\begin{semilogyaxis}[
    width=0.86\linewidth,
    height=3.5cm,
    xlabel={SNR (dB)},
    ylabel={BLER},
    xmin=2,
    xmax=6,
    ymin=1e-7,
    ymax=1.1,
    xtick={2,2.5,3,3.5,4,4.5,5,5.5,6},
    tick label style={font=\footnotesize},
    label style={font=\small},
    legend image post style={scale=0.85},
    xmajorgrids,
    ymajorgrids,
    major grid style={solid,gray!30},
legend style={
    at={(0.5,1.04)},
    anchor=south,
    legend columns=2,
    legend cell align=left,
    draw=gray!60,
    line width=0.4pt,
    fill=white,
    fill opacity=1,
    text opacity=1,
    font=\scriptsize,
    inner sep=2pt,
    row sep=1pt,
    /tikz/every even column/.append style={column sep=5pt}
}
]
\addplot[
    mark=*,
    mark size=2pt,
    color=gray!75!black,
    thick
] coordinates {
(2,1)
(2.5,1)
(3,1)
(3.5,1)
(4,0.999966168388)
(4.5,0.9811981)
(5,0.6281467)
(5.5,0.100831704029)
(6,0.00360282788412)
};
\addlegendentry{LDPC-BP}

\addplot[
    mark=diamond*,
    mark size=2.4pt,
    color=red!85!black,
    thick
] coordinates {
(2,0.0656516335227)
(2.5,0.0109654017857)
(3,0.00119835421495)
(3.5,0.000113079068514)
(4,2.09495752831e-5)
(4.5,6.3e-6)
(5,2.6e-6)
(5.5,1.07191104919e-6)
(6,4.54667140005e-7)
};
\addlegendentry{SF-BP, $N_{\rm sem}=1$}

\addplot[
    mark=square*,
    mark size=2pt,
    color=black!85!black,
    thick
] coordinates {
(2,0.0170010653409)
(2.5,0.00194382440476)
(3,0.000188975593859)
(3.5,3.85570462840e-5)
(4,8.19765989340e-6)
(4.5,3e-6)
(5,1.9e-6)
(5.5,8.24546960918e-7)
(6,4.54667140005e-7)
};
\addlegendentry{SF-BP, $N_{\rm sem}=3$}

\addplot[
    mark=triangle*,
    mark size=2.4pt,
    color=orange!85!black,
    thick
] coordinates {
(2,0.00905539772727)
(2.5,0.000934709821429)
(3,0.000113159038239)
(3.5,2.44008792303e-5)
(4,6.50607928048e-6)
(4.5,2.9e-6)
(5,1.4e-6)
(5.5,8.24546960918e-7)
(6,4.54667140005e-7)
};
\addlegendentry{SF-BP, $N_{\rm sem}=6$}

\end{semilogyaxis}
\end{tikzpicture}
\caption{ BLER versus SNR for different maximum \black{LM}-invocation
budgets for \black{the $(1248,1040)$} rate-$5/6$  LDPC code.}
\label{fig:bler_semantic_budget}
\vspace{-1.3em}
\end{figure}

\subsubsection{Effect of LM-Invocation Budget}
\label{sec:lm_budget}

Fig.~\ref{fig:bler_semantic_budget} evaluates the LM-invocation budget \black{$N_{\rm sem}$}. \black{A single LM invocation} already reduces
\black{the} BLER by several orders of magnitude at moderate SNR. \black{Additional LM invocations}
mainly improve the low-SNR region\black{, where} the
semantic factors \black{are refreshed} from updated text estimates. \black{The small gap between $N_{\rm sem}=3$ and $N_{\rm sem}=6$
at high SNR shows diminishing returns from further invocations.}

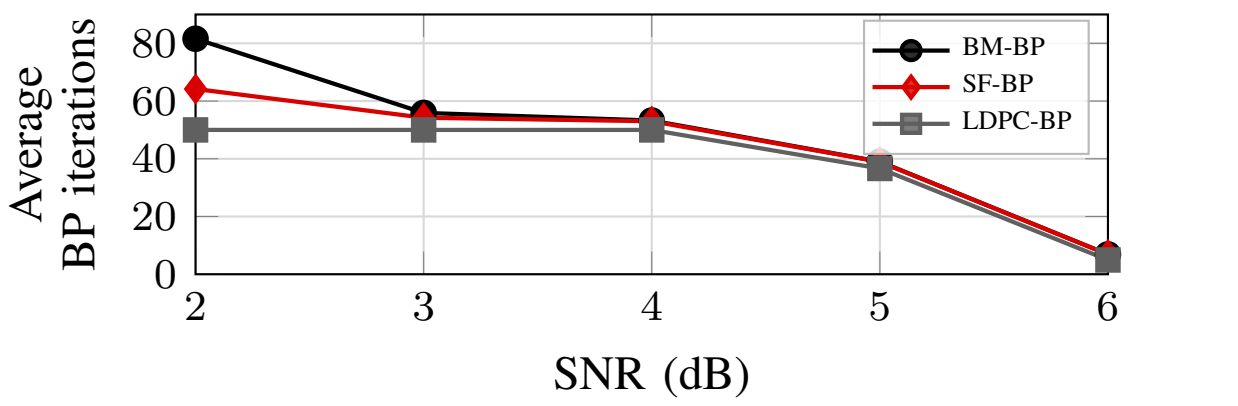
\begin{figure}[t]
\centering
\begin{tikzpicture}
\begin{axis}[
    width=0.86\linewidth,
    height=3.3cm,
    xlabel={SNR (dB)},
    ylabel style={align=center},
    ylabel={Average \\ BP iterations},
    xmin=2,
    xmax=6,
    ymin=0,
    ymax=90,
    xtick={2,3,4,5,6},
    ytick={0,20,40,60,80},
    tick label style={font=\footnotesize},
    label style={font=\small},
    legend image post style={scale=0.85},
    xmajorgrids,
    ymajorgrids,
    major grid style={solid,gray!30},
    legend style={
        at={(0.98,0.98)},
        anchor=north east,
        legend columns=1,
        legend cell align=left,
        draw=gray!55,
        font=\tiny,
        inner sep=1.2pt,
        row sep=-2pt,
        fill=white,
        fill opacity=0.82,
        text opacity=1
    }
]

\addplot[
    mark=*,
    mark size=2.1pt,
    color=black!85!black,
    thick
] coordinates {
(2,81.5836181641)
(3,55.8994330249)
(4,53.2562097755)
(5,39.0047268)
(6,6.88029169293)
};
\addlegendentry{BM-BP}

\addplot[
    mark=diamond*,
    mark size=2.3pt,
    color=red!85!black,
    thick
] coordinates {
(2,64.1882102273)
(3,54.2269913728)
(4,53.0137304297)
(5,38.9423151)
(6,6.87673000847)
};
\addlegendentry{SF-BP}

\addplot[
    mark=square*,
    mark size=2pt,
    color=gray!75!black,
    thick
] coordinates {
(2,50)
(3,50)
(4,49.9991989715)
(5,36.6463294)
(6,4.93633491545)
};
\addlegendentry{LDPC-BP}

\end{axis}
\end{tikzpicture}
\caption{Average total number of BP iterations for the $(1248,1040)$ rate-$5/6$ LDPC code.}
\label{fig:avg_bp_iterations}
\vspace{-1em}
\end{figure}

\subsubsection{Iteration Behavior}
\black{
Fig.~\ref{fig:avg_bp_iterations} reports total BP iterations
over the initial and all  \black{subsequent BP} runs.
Both SF-BP and BM-BP perform at least  two BP runs per block.
As SNR increases, the runs shorten and fewer additional runs
are needed. SF-BP uses fewer iterations than BM-BP.}

\subsubsection{Decoding Latency}
Table~\ref{tab:decoding_latency} reports mean wall-clock
latency per block, measured from channel observations to
final information-bit decisions.
\black{Measurements use an RTX 5090 GPU and an Intel Core Ultra 7
265 CPU with one CPU thread, batch size one, and FP32.}
\black{BM-BP and SF-BP allow up to $N_{\rm sem}=6$ LM invocations
with syndrome-based early termination; LM-Top1 uses at most one.
Frequent semantic processing increases latency at low SNR.
SF-BP is slower than BM-BP because it recomputes
semantic messages at each BP iteration; fewer LM invocations
partly offset this cost at low SNR.
At $6$~dB, BM-BP and SF-BP each use one LM invocation per
block and remain slower than LDPC-BP.}

\begin{table}[t]
\tiny
\centering
\color{black}
\caption{Average End-to-End Decoding Latency (ms/block)}
\label{tab:decoding_latency}
\renewcommand{\arraystretch}{0.9}
\resizebox{\columnwidth}{!}{%
\begin{tabular}{c cccc cc}
\toprule
& \multicolumn{4}{c}{Mean latency (ms/block)}
& \multicolumn{2}{c}{\black{Mean LM invocations/block}} \\
\cmidrule(lr){2-5}\cmidrule(lr){6-7}
SNR (dB)
& LDPC-BP & LM-Top1 & BM-BP & SF-BP
& BM-BP & SF-BP \\
\midrule
2.0 & \(3.874\) & \(9.283\)
& \(15.637\) & \(24.167\)
& 1.40425 & 1.13275 \\
3.0 & \(3.831\) & \(9.293\)
& \(10.897\) & \(13.541\)
& 1.01625 & 1.00050 \\
4.0 & \(3.913\) & \(9.399\)
& \(10.638\) & \(12.523\)
& 1.00025 & 1.00000 \\
5.0 & \(3.062\) & \(6.426\)
& \(9.729\) & \(11.009\)
& 1.00000 & 1.00000 \\
6.0 & \(0.525\) & \(0.542\)
& \(7.582\) & \(8.525\)
& 1.00000 & 1.00000 \\
\bottomrule
\end{tabular}%
}
\end{table}

\section{Conclusion}

This paper introduced a semantic Tanner graph that represents each
\black{LM} token posterior as a joint factor over \black{the bits of the token}. SF-BP performs  message passing on the augmented \black{Tanner}
graph, \black{where the semantic factors} and \black{the} LDPC parity checks
interact through shared \black{variable nodes}. \black{Simulations \black{with a ByT5-small model fine-tuned for each code configuration} showed that SF-BP lowers the BLER by several orders of magnitude compared with BP and outperforms other LM-aided BP receivers in both BLER and text quality.}

\bibliographystyle{IEEEtran}
\bibliography{ref_letter}
\end{document}